\documentclass[a4paper,11pt]{article}
\pdfoutput=1
\usepackage{jheppub} 
\usepackage{natbib}
\setcitestyle{square,comma,numbers,sort&compress}
\usepackage{enumerate}
\usepackage{hyperref}
\usepackage{tabularx,booktabs}
\usepackage[dvipsnames]{xcolor}
\usepackage{comment}
\usepackage[caption=false]{subfig}
\usepackage{cancel}
\usepackage{fontawesome}
\usepackage{hyperref}

\usepackage[section]{placeins}

\newcolumntype{Y}{>{\centering\arraybackslash}X}

\definecolor{myRED}{rgb}{0.8, 0.25, 0.33}

\newcommand{\rmv}[1]{{\iffalse #1 \fi}}

\usepackage[force]{feynmp-auto}
\usepackage[T1]{fontenc} 

\usepackage{cleveref}

\usepackage{pifont}
\usepackage{mathtools}

\usepackage{feynmp-auto}

\def\be{\begin{equation}}
\def\ee{\end{equation}}

\def\bea{\begin{eqnarray}}
\def\eea{\end{eqnarray}}

\def\be{\begin{equation}}
\def\ee{\end{equation}}

\begin{document}

\title{Minimal Multi-Majoron Model}

\author[a,b]{Bowen Fu,}
\author[c]{Anish Ghoshal}
\author[d]{and Stephen F. King}
\affiliation[a]{Key Laboratory of Cosmology and Astrophysics (Liaoning) \& College of Sciences, Northeastern University, Shenyang 110819, China}
\affiliation[b]{Foshan Graduate School of Innovation, Northeastern University, Foshan 528312, China}
\affiliation[c]{Institute of Theoretical Physics, Faculty of Physics, University of Warsaw, ul. Pasteura 5, 02-093
Warsaw, Poland}
\affiliation[d]{School of Physics \& Astronomy, University of Southampton, Southampton SO17 1BJ, UK}

\emailAdd{fubowen@neu.edu.cn}
\emailAdd{anish.ghoshal@fuw.edu.pl}
\emailAdd{king@soton.ac.uk}

\abstract{In order to provide a natural framework for hierarchical right-handed neutrinos, we propose a realistic ultraviolet complete minimal multi-Majoron model (MMMM). We consider two right-handed neutrinos for simplicity, although the model is readily extendable to more.
The minimal model introduces two complex scalar Majoron fields $\phi_1$ and $\phi_2$, whose couplings to the two respective right-handed neutrinos are controlled by an extra global $U(1)_N$ symmetry.  
We show that a flavon field is required to facilitate the effective Yukawa couplings, in order to implement the type I seesaw mechanism.
We analyse the resulting phenomenology related to neutrino masses, flavour mixing and cosmological predictions concerning the formation and decay of topological defects like the global cosmic strings and the domain walls when the $U(1)_N\times U(1)_{B-L}$ symmetry is broken. The resulting gravitational wave spectrum is a distinctive combination of the spectrum from the global cosmic string and strong first-order phase transitions when the symmetries are broken, 
the strength of the latter being enhanced by the second Majoron field.
The resulting characteristic spectrum determines the two right-handed neutrino mass scales within the considered framework. 
}

\maketitle
\flushbottom

\section{Introduction}

Neutrino mass and mixing is the most concrete evidence of physics beyond the standard model in the present day \cite{Super-Kamiokande:1998kpq,Super-Kamiokande:2001ljr,DayaBay:2012fng}. 
One of the most popular explanations is the type I seesaw mechanism \cite{Minkowski:1977sc,Yanagida:1979as,GellMann:1980vs,Mohapatra:1979ia}, where the left-handed neutrinos obtain their mass through mixing with a set of heavy right-handed neutrinos (RHNs).
Through the oscillation experiments, the mass square splittings of the left-handed neutrino mass have been measured as $\Delta m_{21}^2 \simeq 7.5 \times 10^{-5} \mathrm{eV}^2 $ and $\Delta m_{3\ell}^2 \simeq 2.5 \times 10^{-3} \mathrm{eV}^2 $ \cite{Esteban:2024eli}, while the cosmological bound for the sum of left-handed neutrino masses is approaching $0.1$ eV \cite{Palanque-Delabrouille:2019iyz, eBOSS:2020yzd}. 
The combined result indicates the left-handed neutrino mass spectrum is likely to be hierarchical. 

The hierarchy in fermion masses has been recognised as an unexplained puzzle for a long time \cite{Froggatt:1978nt,Barr:1979xt,Barbieri:1983uq,Balakrishna:1987qd,Babu:1988fn,Babu:2004th,Babu:2009fd} and has drawn a great deal of attention~\cite{Babu:1999me,Yoshioka:2000tve,Chao:2012re,Altmannshofer:2014qha,Higuchi:2014cua,Huitu:2017ukq,Okada:2019fgm,Hernandez:2021iss,Bonilla:2023wok,FernandezNavarro:2024hnv,Arbelaez:2024rbm,Jana:2024icm}. 
The seesaw mechanism can help explain the smallness of neutrino mass, but cannot shed light on the origin of the charged fermion mass hierarchy.  
In many models, such as $SO(10)$ the right-handed neutrinos involved in the type I seesaw mechanism are also expected to have a strong hierarchical mass spectrum (for a recent example see e.g.~\cite{Babu:2024ahk}).
A potential explanation for hierarchical right-handed neutrino masses are multi-Majoron models \cite{Bamert:1994hb}, where different right-handed neutrinos $N_i$ couple to their own respective (personal) complex scalar Majoron fields $\phi_i$ through Yukawa interaction
\begin{eqnarray}
  \frac{y^N_i}{2}\phi_i \overline{N_i^c}N_i + h.c.
\end{eqnarray} 
with coupling constant $y^N_i$. 
Then the hierarchical mass spectrum of right-handed neutrinos can be explained by the different scales at which the complex scalar Majoron fields get vacuum expectation values (VEVs).

Recently, the gravitational wave (GW) from phase transitions in a multi-Majoron model has been studied \cite{DiBari:2023mwu}. Due to the coupling between the new scalar fields and the hierarchical symmetry breaking pattern, the strength of the first-order phase transition can be significantly enhanced, which leads to testable signatures in upcoming GW detectors. 
In the meantime, the global cosmic strings provide an additional source of GW, making the total signal distinctive. \footnote{A recent analysis shows interesting compatibility of such a signal with the observed NANOGrav result based on a Majoron interpretation \cite{Ghosh:2025cxp}.} 
However, the studied scenario~\cite{DiBari:2023mwu} did not include an explicit model: such a model would need to ensure that each right-handed neutrino couples to different heavy scalars. In order to achieve this, the right-handed neutrinos $N_i$ must be distinguishable, along with the complex scalar Majoron fields $\phi_i$, from which the Majoron Goldstone bosons emerge. 
Consequently, each lepton doublet can only couple to one right-handed neutrino, and the resulting SM neutrino mass matrix is diagonal (or block-diagonal).
\footnote{For example, in the minimal seesaw model with two RHNs, if $N_1$ couples to $L_e$ and $N_2$ couples to $L_{\mu,\tau}$, the light neutrino mass matrix given by the type-I seesaw mechanism would be block-diagonal in the form of  
\begin{eqnarray*}
    \begin{pmatrix}
        \times & 0 & 0 \\
        0 & \times & \times \\
        0 & \times & \times
    \end{pmatrix}\,,
\end{eqnarray*}
which only allows the mixing between two lepton flavours. 
The situation cannot be improved by charged lepton mixing.}
Consequently, the lepton mixing cannot be properly reproduced. The main purpose of this paper is to provide a consistent multi-Majoron model and investigate its consequences.

In this paper, then, we present the first realistic multi-Majoron model which provides a natural setting for hierarchical right-handed neutrinos.
For simplicity, we focus on a minimal version with just two right-handed neutrinos and two Majoron fields: the minimal multi-Majoron model (MMMM). However the model may be readily extended to more right-handed neutrinos and corresponding Majorons. 
The two complex scalar Majoron fields $\phi_1$ and $\phi_2$, couple to the two respective right-handed neutrinos due to an extra global $U(1)_N$ symmetry, with a flavon field being necessary to obtain the effective Yukawa couplings, and so implement the type I seesaw mechanism. The model predicts global cosmic strings (and domain walls) when the $U(1)_N\times U(1)_{B-L}$ symmetry is broken. We discuss the resulting gravitational wave spectrum from the global cosmic string and also the strong first-order phase transitions (FOPT) when the symmetries are broken, 
the strength of the latter being enhanced by the second Majoron field.
The resulting characteristic spectrum determines the two right-handed neutrino mass scales within the considered framework.

\textit{The paper is organised as follows:} 
In Sec.\ref{sec:model}, we present the minimal multi-Majoron model as a concrete example of realistic multi-Majoron models. 
In Sec.\ref{sec:pheno}, we discuss the possible phenomenological consequences of the model, including neutrino mass and mixing, GW from FOPT and cosmic string.
In Sec.\ref{sec:con}, we summarise and conclude.

\section{Minimal Multi-Majoron model \label{sec:model}}

In multi-Majoron models, the RHNs are each coupled to their own personal complex scalar Majoron field.
The hierarchical VEVs of the complex scalar Majoron fields are then responsible for generating the hierarchical masses of the RHNs.
In general, there can be three complex scalar Majoron fields in the seesaw model with three RHNs, each of which couples to one of the RHNs. 
A minimal version of multi-Majoron models can be obtained by reducing the number of RHNs and complex scalar Majoron fields to two.
In any case, there has to be an extra symmetry under which different RHNs are charged so that the coupling between each RHN and its personal respective complex scalar Majoron field can be ensured. 

To illustrate the model explicitly, we present a minimal implementation of the two-Majoron model as a concrete example.
In this model, the SM symmetry is extended by a $U(1)_{B-L}$ symmetry and a $U(1)_N$ symmetry, both of which are global. 
The $U(1)$ charges of the new particles are listed in Tab.~\ref{tab:1flavon}.
\begin{table}[t!]
    \centering
    \begin{tabular}{c|cccc|ccccccccccc}
    & $N^c_{1}$ & $N^c_2$ & $\phi_1$ & $\phi_2$ & $\theta$ & $(\chi_i)_L$ & $(\chi_i)_R^c$\\\midrule
    $U(1)_{B-L}$ & $1$ & $1$ & $2$ & $2$ & $0$ & $-1$ & $1$    \\
    $U(1)_N$ & $-1$ & $1$ & $-2$ & $2$ & $1$ & $0$ & $0$ 
    \end{tabular}
    \caption{\it Charges of the new particles in the global $U(1)_{B-L}\times U(1)_N$ extension of the Standard Model with one flavon field $\theta$. 
    $\phi_1$ and $\phi_2$ are the complex scalar Majoron fields responsible for generating the respective RHN masses.}
    \label{tab:1flavon}
\end{table}
Under the $U(1)_N$ symmetry, the SM particles are neutral while the (CP-conjugated) RHNs $N^c_1$ and $N^c_2$ are charged oppositely. 
Besides the SM particles and the RHNs, there are two scalars $\phi_i$, a scalar $\theta$, and two vector-like fermions $\chi_i$ with $i=1,2$. 
Adopting the Weyl representation for the fermions, the allowed interactions involving the new particles can be summarised as  
\begin{eqnarray}
    \mathcal{L} &\supset& M_{\chi_i} \chi_i^c\chi_i +  y^{e}_{\alpha \beta}H e_\beta^c L_\alpha +  Y^{L}_{\alpha i} H \chi_i^c L_\alpha + Y^{N}_{1i} \theta N_1^c \chi_i + Y^{N}_{2i} \theta^* N_2^c \chi_i \\\nonumber
    &&  + \frac{y^N_1}{2}\phi_1 N^c_1N^c_1 + \frac{y^N_2}{2}\phi_2 N^c_2 N^c_2 + \text{h.c.}\,,
\end{eqnarray}
where $L_\alpha$ are the SM lepton doublets, $e_\beta$ are the charged leptons and $H$ is the Higgs doublet. 
$y^e_{\alpha \beta}$, $Y^{L}_{\alpha i}$, $Y^{N}_{ij}$ and $y^N_i$ are the Yukawa couplings and $M_{\chi_i}$ is the mass of $\chi_i$. 
\footnote{In general, there can be a mass mixing between $\chi_1$ and $\chi_2$, but there always exists a basis where such mixing disappears.}

\begin{figure}
\centering
\begin{fmffile}{dirac_seesaw}
\fmfframe(10,20)(10,10){
\begin{fmfgraph*}(180,45)
\fmflabel{$N_{1,2}^c$}{o1}
\fmflabel{}{o2}
\fmflabel{}{i2}
\fmflabel{$L_\alpha$}{i1}
\fmfv{}{v1}
\fmfv{decor.shape=cross,decor.size=8,label=$M_{\chi}$,label.angle=90}{v2}
\fmfv{}{v3}
\fmfv{label=$H$,label.angle=90}{v4}
\fmfv{}{v5}
\fmfv{label=$\theta,,\theta^*$,label.angle=90}{v6}
\fmfleft{i1,i2}
\fmfright{o1,o2}
\fmf{fermion}{i1,v1}
\fmf{fermion,label=$\chi_i$,l.side=right}{v1,v2}
\fmf{fermion,label=$\chi_i$,l.side=right}{v2,v3}
\fmf{fermion}{v3,o1}
\fmf{phantom}{i2,v4}
\fmf{phantom}{v5,v4}
\fmf{phantom}{v5,v6}
\fmf{phantom}{o2,v6}
\fmf{dashes,tension=0}{v1,v4}
\fmf{phantom,tension=0}{v2,v5}
\fmf{dashes,tension=0}{v3,v6}
\end{fmfgraph*}}
\end{fmffile}
\caption{\it Feynman diagram responsible for generating effective Yukawa couplings (represented by a hashed blob in the next figure). \label{fig:dirac_seesaw}}
\end{figure}
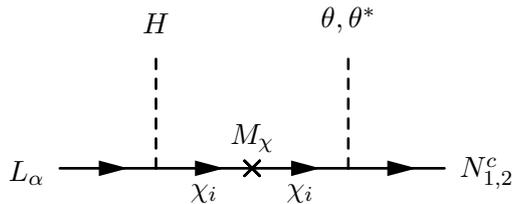

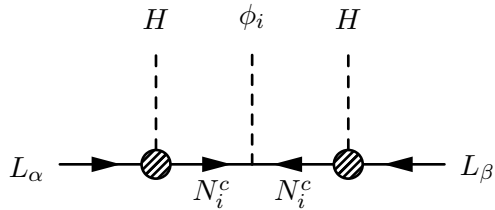
\begin{figure}
\centering
\begin{fmffile}{type_I}
\fmfframe(10,20)(10,10){
\begin{fmfgraph*}(180,45)
\fmflabel{$L_\beta$}{o1}
\fmflabel{}{o2}
\fmflabel{}{i2}
\fmflabel{$L_\alpha$}{i1}
\fmfblob{.06w}{v1}
\fmfv{}{v2}
\fmfblob{.06w}{v3}
\fmfv{label=$H$,label.angle=90}{v4}
\fmfv{label=$\phi_i$,label.angle=90}{v5}
\fmfv{label=$H$,label.angle=90}{v6}
\fmfleft{i1,i2}
\fmfright{o1,o2}
\fmf{fermion}{i1,v1}
\fmf{fermion,label=$N_i^c$,l.side=right}{v1,v2}
\fmf{fermion,label=$N_i^c$,l.side=left}{v3,v2}
\fmf{fermion}{o1,v3}
\fmf{phantom}{i2,v4}
\fmf{phantom}{v5,v4}
\fmf{phantom}{v5,v6}
\fmf{phantom}{o2,v6}
\fmf{dashes,tension=0}{v1,v4}
\fmf{dashes,tension=0}{v2,v5}
\fmf{dashes,tension=0}{v3,v6}
\end{fmfgraph*}}
\end{fmffile}\label{fig:typeI_seesaw}
\caption{\it Feynman diagram for the effective Weinberg operators due to the type I seesaw mechanism with Majorons. Note the diagonal coupling of the $i-th$ right-handed neutrino $N^c_i$ to its personal Majoron field $\phi_i$.}
\end{figure}

Because of the $U(1)$ charges, $N^c_1$ can only couple to $\phi_1$ and $N^c_2$ can only couple to $\phi_2$, making the origin of the Majorana masses of the two RHNs different. 
Meanwhile, the Yukawa interactions $HLN^c_i$ are forbidden. 
Nevertheless, $\theta$ through dim-5 interactions $H\theta N_1^c L_\alpha$ and $H\theta^* N_2^c L_\alpha$
can be generated through the Dirac seesaw mechanism as $\chi$ is integrated out. 
The corresponding Feynman diagram is shown in Fig.\ref{fig:dirac_seesaw}.
After the flavon field $\theta$ gets VEVs $\langle\theta\rangle$, Yukawa interactions can be generated by the Dirac seesaw mechanism, which read
\begin{eqnarray}
    Y^{\nu}_{\alpha 1} H N_1^c L_\alpha 
    + Y^{\nu}_{\alpha 2} H N_2^c L_\alpha + \text{h.c.}\,, 
    \quad \text{with } 
    Y^{\nu}_{\alpha i} = \sum_{j=1,2}Y^{L}_{\alpha j} Y^{N}_{ij} \frac{\langle\theta\rangle}{M_{\chi_j}}\,. \label{eq:Yukawa_eff}
\end{eqnarray}
Then the SM neutrino mass can be obtained by the standard type I seesaw mechanism as shown in Fig.\ref{fig:typeI_seesaw}.
The RHNs obtain their masses from the VEV of $\phi_i$.

\subsection{Surviving symmetries}

In the spirit of the seesaw mechanism, we assume the VEV of the flavon field $\langle \theta \rangle$ is small compared with $\langle \phi_i \rangle$ and $M_\chi$.
In that case, the $U(1)_N$ and $U(1)_{B-L}$ are broken by $\langle \phi_i \rangle$ at two different scales.
\footnote{Note that the $U(1)_{B-L}$ here can only be a global symmetry since there are only two right-handed neutrinos. 
However, the model can readily be extended to include a third right-handed neutrino with a gauged $U(1)_{B-L}$ symmetry if necessary.}
If $\langle \phi_1 \rangle < \langle \phi_2 \rangle$, the $U(1)_N$ and $U(1)_{B-L}$ are both broken by $\langle \phi_2 \rangle$ first. 
However, the VEV of $\phi_2$ can only break one $U(1)$ symmetry, under which the charges of the particles are the sum of their $U(1)_N$ charge and $U(1)_{B-L}$ charge, namely $U(1)_{N+(B-L)}$, as shown in Tab.~\ref{tab:1flavon_bc2}. 
\begin{table}
    \centering
    \begin{tabular}{c|ccccccccccccccc}
    & $L_{\alpha}$ & $e^c_{\alpha}$  & $N^c_{1}$ & $N^c_2$ & $\phi_1$ & $\phi_2$ & $\theta$ & $(\chi_i)_L$ & $(\chi_i)_R^c$\\\midrule
    $U(1)_{N-(B-L)}$  & $1$ & $-1$ & $-2$ & $0$ & $-4$ & $0$ & $1$ & $1$ & $-1$   \\
    $U(1)_{N+(B-L)}$  & $-1$ & $1$ & $0$ & $2$ & $0$ & $4$ & $1$ & $-1$ & $1$   \\ \midrule
    $Z_4^I (< U(1)_{N-(B-L)})$ & $i$ & $-i$ & $-1$ & $1$ & $1$ & $1$ & $i$ & $i$ & $-i$  \\ 
    $Z_4^{II} (< U(1)_{N+(B-L)})$ & $-i$ & $i$ & $1$ & $-1$ & $1$ & $1$ & $i$ & $-i$ & $i$ 
    \end{tabular}
    \caption{\it Charges of the particles under the recombined $U(1)$ of $U(1)_{B-L}$ and $U(1)_N$ and residue discrete symmetries after the $U(1)$'s are broken.}
    \label{tab:1flavon_bc2}
\end{table}
The other orthogonal linear combination of $U(1)_N$ and $U(1)_{B-L}$, namely $U(1)_{N-(B-L)}$, remains unbroken after $\phi_2$ gets its VEV. 
The symmetry $U(1)_{N+(B-L)}$ is broken into a $Z_4^{II}$ symmetry under which all the fermions and $\theta$ are charged. 
Then, when $\phi_1$ gains a VEV $\langle \phi_1 \rangle$, the $U(1)_{N-(B-L)}$ is broken into another $Z_4^{I}$ symmetry under which only $N_i$ and $\theta$ are charged. 
The $Z_4^{I}$ and $Z_4^{II}$ symmetries are both broken by the VEV of $\theta$ at a lower scale. 
As $N_1$ and $N_2$ are charged differently under the $Z_4^{I}$ and $Z_4^{II}$ symmetries, the $Z_4^{I}$ and $Z_4^{II}$ symmetries can also be recognised as flavour symmetries, with $\theta$ the flavon field. 
The breaking chain can be thus summarised as
    \begin{eqnarray}
    &&U(1)_{N-(B-L)}
    \stackrel{\langle \phi_1 \rangle}{\xrightarrow{\hspace{0.8cm}}} 
    Z_4^{I} \stackrel{\langle \theta \rangle}{\xrightarrow{\hspace{0.8cm}}}
    I \,,
    \\
    &&U(1)_{N+(B-L)}
    \stackrel{\langle \phi_2 \rangle}{\xrightarrow{\hspace{0.8cm}}} 
    Z_4^{II} \stackrel{\langle \theta \rangle}{\xrightarrow{\hspace{0.8cm}}}
    I \,.
    \label{eq:bc_1}
    \end{eqnarray}
If $\langle \phi_1 \rangle > \langle \phi_2 \rangle$, then the $U(1)_{N-(B-L)}$ is broken first by the VEV of $\phi_1$, and the $U(1)_{N+(B-L)}$ survives until $\phi_2$ gains a VEV. 

When $Z_4^{I}$ and $Z_4^{II}$ are broken, cosmic domain walls can be generated. 
If there is a cosmic string relic from $U(1)$ breaking at the time of wall formation, the domain walls would be bounded by the strings, forming hybrid topological defects that are named as walls bounded by strings \cite{Kibble:1982dd, Everett:1982nm, Lazarides:2023iim}. 

The symmetry breaking also results in multiple pseudo-Nambu-Goldstone bosons (pNGBs). 
Apart from the pNGBs left from $\phi_1$ and $\phi_2$, which are recognised as the Majorons, there is also one left from $\theta$.
These pNGBs can be related to many phenomena, including  dark matter \cite{Berezinsky:1993fm,Gu:2010ys,Garcia-Cely:2017oco,Ma:2017xxj,Akita:2023qiz,King:2024idj} and leptogenensis \cite{Vilja:1993uw,Pilaftsis:2008qt,Ibe:2015nfa,Brune:2022vzd,Wada:2024cbe,Brune:2025zwg}.
The mass spectrum of these pseudo-Goldstone bosons would be determined by the soft symmetry breaking terms, like $(\phi_i^2 + {\phi_i^*}^2)$.

\medskip

\section{Phenomenology of MMMM \label{sec:pheno}}

\subsection{Neutrino mass and mixing}

After all scalars get VEVs, the Dirac mass terms of $\nu$, $N$, and $\chi_{L,R}$ can be written in matrix form as
\begin{eqnarray}
    \begin{pmatrix}
        N^c & \chi_R^c
    \end{pmatrix}
    \begin{pmatrix}
        0 & Y^N\langle\theta\rangle\\
        (Y^L)^T\langle H\rangle & M_{\chi}
    \end{pmatrix}
    \begin{pmatrix}
        \nu \\ \chi_L
    \end{pmatrix}
\end{eqnarray}
where $(Y^{L})_{\alpha i} = Y^{L}_{\alpha i}$, $(Y^{N})_{ai}=Y^{N}_{ai}$ and $(M_\chi)_{ij}=\delta_{ij}M_{\chi_i}$. 
Through the Dirac seesaw mechanism, the neutrino Dirac mass term $M_DN^c\nu$ can be generated, with $M_D$ given by 
\begin{eqnarray}
    M_D = Y^{N}\langle\theta\rangle  M_\chi^{-1} (Y^L)^T \langle H\rangle\,.
\end{eqnarray}
Then the normal type I seesaw mass matrix for the neutrinos is recovered, which reads
\begin{eqnarray}
    \frac12 \begin{pmatrix}
        \nu & N^c
    \end{pmatrix}
    \begin{pmatrix}
        0 & M_D\\
        M_D^T & M_N
    \end{pmatrix}
    \begin{pmatrix}
        \nu \\ N^c
    \end{pmatrix} + \text{h.c.}
\end{eqnarray}
where $(M_N)_{ij} = \delta_{ij}y^N_i\langle\phi_i\rangle $.
The light neutrino mass matrix is raised as 
\begin{eqnarray}
    \Big(m_\nu\Big)_{\alpha\beta} = M_D^T M_N^{-1}M_D\,.
\end{eqnarray}
Since $Y^L_{\alpha 1}$ and $Y^L_{\alpha 2}$ are linear independent and can both be non-zero for all 3 lepton flavours, the neutrino mixing can be reproduced easily.
At scale below $\langle\theta\rangle$, the model is effective the same as the type-I seesaw model with two right-handed neutrinos \cite{King:1999mb,Ibarra:2003up}, which is known to be able to explain the neutrino oscillation completely.

\subsection{Topological defects}

In the breaking chain $U(1)\longrightarrow Z_n \longrightarrow I$, the $U(1)$ symmetry breaking leads to the formation of cosmic strings while the $Z_n$ symmetry breaking can lead to domain wall formation. The hybrid network of the strings and walls in this scenario is dubbed as \textit{wall bounded by strings}.
\footnote{Such a hybrid defect in this scenario is different from the hybrid defect in the axion (a) models \cite{Vilenkin:1982ks}. 
In the axion models, the strings wind from 0 to $2\pi$ as the $U(1)$ symmetry is fully broken when the strings form. 
Later on, when the domain walls form at low scale and the $\cos (Na/f_a)$ potential becomes important, each string can be connected to $N$ walls, where $f_a$ represents the axion decay constant. 
On the other hand, in the scenario of breaking chain $U(1)\longrightarrow Z_n \longrightarrow I$, the strings wind from 0 to $2\pi/n$, as the $U(1)$ symmetry is not fully broken but into $Z_n$ when the strings form. 
Consequently, each string is only connected to one wall after the $Z_n$ is broken.}

The motion of wall bounded by string is determined by the competition between the wall tension and string tension. We follow the paradigm discussed in Ref.\cite{Dunsky:2021tih,Fu:2024rsm} for the case with local strings and provide an estimate for our scenario where global strings are involved. In the case of walls bounded by gauged strings, there are two phases during the evolution of the network:
\begin{itemize}
    \item When the string tension dominates over the wall tension, the behaviour of the hybrid defect is nearly the same as pure global string. 
    \item When the wall tension dominates over the string tension, the hybrid defects collapse very fast, like a transient source of GW, leaving a $f^3$-spectrum \cite{Caprini:2009fx,Cai:2019cdl,Brzeminski:2022haa}. 
\end{itemize}
However, for the walls bounded by global strings, the Goldstone boson emission can also affect the motion of the hybrid defects. 
In the following discussion, we determine the dynamics of the hybrid defect by comparing the power of Goldstone boson emission and GW emission.

We start with a brief review of the properties of the global strings.
The tension of a global string is given by \cite{Chang:2021afa,Fu:2023nrn}
\begin{eqnarray}
    \mu \simeq 2\pi \eta^2 \ln \frac{L}{\delta}\,,
\end{eqnarray}
where $\eta$ is the VEV of the scalar field.
$\delta$ is the radius of the string core, which is roughly the inverse of the mass of the scalar field, or $\eta$.
$L$ is the mean separation between the strings, which is approximately given by $L \sim t$ in the scaling regime.
The length of the string loops at time $t$ is 
\begin{eqnarray}
    l(t)=\alpha t_i-\Gamma_\mathrm{GW} G \mu\left(t-t_i\right)-\kappa_\mathrm{GB}\left(t-t_i\right)\,,
\end{eqnarray}
where $\alpha\simeq 0.1$ characterises the ratio between the loop size and Hubble horizon size at formation, $t_i$ is the time of loop formation and 
\begin{eqnarray}
    \kappa_\mathrm{GB} \equiv \frac{\Gamma_\mathrm{GB}}{2\pi\ln(\eta t)}
\end{eqnarray}
with $\Gamma_\mathrm{GB}\approx 65$ the production rate of Goldstone bosons. 
$\kappa_\mathrm{GB}$ is typically much larger than $\alpha$ and thus the global strings decay very fast after formation.

Once the walls form, the hybrid defect would have different GW emission rates for different sizes of loops. 
The wall tension is approximately given by $\mathcal{E}\sim \langle \theta \rangle^3$. 
The string and the wall can both drive the GW emission: if $R < R_c \equiv \mu/\mathcal{E}$, the strings drive the GW emission and the power is similar to the pure string case, i.e. $P_\mathrm{GW} = \Gamma_\mathrm{GW} G \mu^2$; if $R > R_c$, the domain wall tension drive the GW emission and the power is enhanced by $(R/R_c)^2$, becoming $P_\mathrm{GW} = \Gamma_\mathrm{GW} G \mathcal{E}^2 R^2$.
However, the GW emission can dominate over the Goldstone boson emission only when $P_\mathrm{GW} /\mu > \kappa$, 
which leads to the condition 
\begin{eqnarray}
    R > \sqrt{\frac{\Gamma_\mathrm{GB}\eta^2}{\Gamma_\mathrm{GW}G\mathcal{E}^2}}\simeq \frac{M_\mathrm{pl} \eta}{\langle \theta \rangle^3} \simeq 30 \frac{\eta}{\langle \theta \rangle} \left(\frac{T}{\langle \theta \rangle}\right)^2 t\,,\label{eq:wall_condition}
\end{eqnarray}
where the relation between temperature and time is assumed to be the one in the radiation dominating era, i.e. $t \simeq 0.03 M_\mathrm{pl}/T^2$.  
As the string loop is short-lived, the temperature $T$ should be close to the scale of wall formation $\langle \theta \rangle$, i.e. $T/\langle \theta \rangle\sim 1$. 
The emission time $t$ is also around the loop formation time $t_i$.
On the other hand, the radius of the string loop at formation is related to the loop formation time by $R=l/2\pi \simeq 0.02\, t_i$, which is in contradiction with Eq.\eqref{eq:wall_condition}, providing the ordering the symmetry breaking order $\eta>\langle \theta \rangle$.
Therefore, the wall cannot dominate the motion of the hybrid defect, and the GW emitted by the hybrid defect is indistinguishable from that emitted by the pure global strings.

\begin{figure}
    \centering
    \subfloat[string-wall hybrid]{\includegraphics[width=0.23\linewidth]{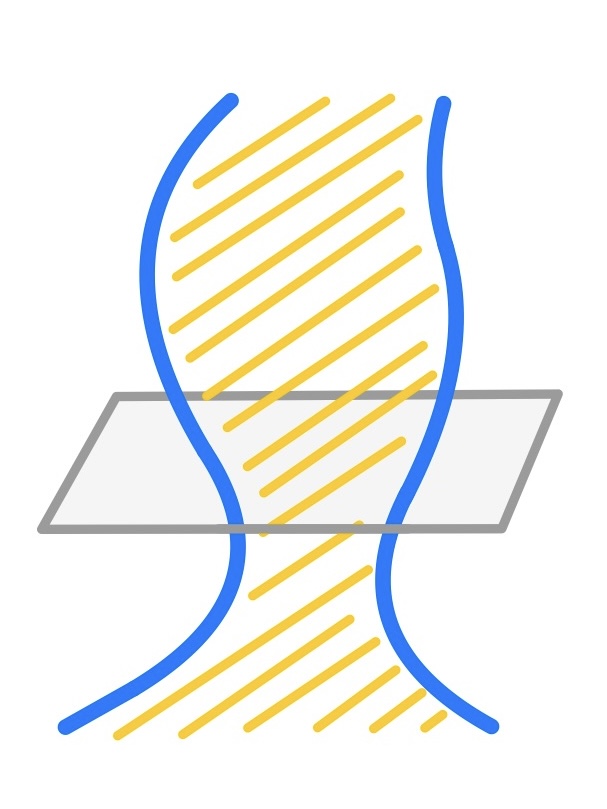}\label{fig:string-wall}}
    \subfloat[cross section of the hybrid defect and the $U(1)$ phases around]{\includegraphics[width=0.75\linewidth]{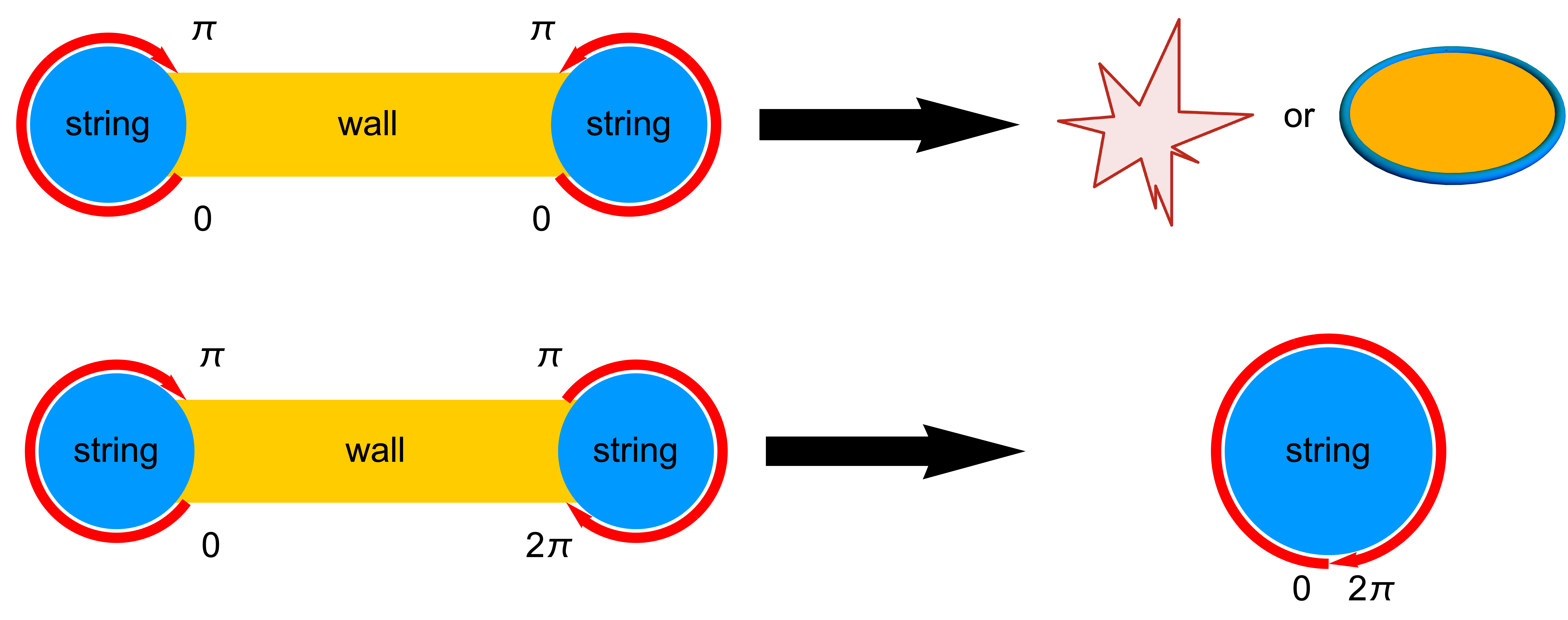}\label{fig:cross-section}}
    \caption{\it Sketch of the hybrid defect formed by long strings and walls. 
    (a) two long strings (blue lines) connected by a wall (yellow region). 
    (b) the cross section of the hybrid defect along the gray plane in (a), with $U(1)$ phases marked out. 
    The thickness of the walls and the radius of the strings are only illustrative and do not represent their real values. }
    \label{fig:string-wall_sketch}
\end{figure}

Although the domain wall cannot dominate the motion of string loops, it can still affect the fate of the long strings.
In particular, two long strings connected by a wall sheet would be accelerated toward each other due to the wall tension. 
In order to illustrate the process in detail, we consider two long strings connected by a wall in a toy model where a $U(1)$ symmetry is broken into a $Z_2$ symmetry and then nothing, i.e. $U(1) \rightarrow Z_2 \rightarrow I$.
The hybrid defect is shown in Fig.\ref{fig:string-wall}, where the blue lines stand for the strings and the yellow region in between stands for the wall.
For the convenience of illustration, the cross section of the hybrid defect along the grey plane is also shown in Fig.\ref{fig:cross-section}. 
As the strings are formed from the symmetry breaking $U(1) \rightarrow Z_2$, $0$ and $\pi$ represent the charges of the unbroken $Z_2$ and the minimal winding angle of the strings is $\pi$. 
Depending on the winding direction of the strings connected by the wall, there are two distinctive processes that can happen:
\begin{itemize}
\item When two strings winding in opposite directions are connected by a wall, the hybrid object has no topological charge. The strings would be pulled together by the wall, annihilating or forming string loops with walls inside, as shown in the upper panel of Fig.\ref{fig:cross-section}.
\item When two strings winding in the same direction are connected by a wall, the hybrid object has a non-trivial topological charge. When the strings are pulled together by the wall, they will merge into a new string with twice the winding angle, as shown in the lower panel of Fig.\ref{fig:cross-section}.
\end{itemize}
In either case, we assume that the process is instantaneous. 
A naive estimation of the probability for each process to happen is 50\%. 
In the first case, the string number density is reduced to half as two strings are merged into one. 
In the second case, the strings simply annihilate or intersect into walls bounded by string loops, which decay shortly after formation. 
Therefore, the total string number is reduced to $1/4$ of that before the wall appears. 
However, the reduced string number also leads to a lower loop formation rate, and the string network can return to the scaling regime after some evolution.
In the meantime, the string tension is doubled as $\mu \propto n$ with $n$ the winding number \cite{Vilenkin:2000jqa}.
As a result, we naively expect the GW produced by the string network to be enhanced by a factor of 4 as $\Omega_\mathrm{gw} \propto \mu^2$. 

In the minimal multi-Majoron model, the naive expectation can be summarised as the following stages:
\begin{itemize}
    \item At $T\sim \langle\phi_i\rangle$, a U(1) symmetry is broken and global cosmic strings with tension $\mu$ are formed. The string network evolves normally and emits both Goldstone bosons and GWs.
    \item At $T\sim \langle\theta\rangle$, the walls form and join the pre-existing strings into a string-wall network. The walls promote the intersection of strings, leading to a higher rate of loop formation and thus a peak in GW production.
    \item As the string-wall network evolves, the walls disappear and there left a pure string network again. Compared with the string network before the wall formation, with tension of the strings is doubled to $2\mu$, but the long string density is reduced to a quarter. 
    \item Shortly later, the long string density grows back to the scaling regime, and the string network evolves as usual. Compared with the string network before wall formation, the string tension is doubled, and thus the GW emitted is 4 times stronger.  
\end{itemize}
The existing time of the wall and how long it takes for the string network to get back to the scaling regime need to be determined by numerical simulations, which is beyond the scope of this work.
On the other hand, the residual discrete symmetry of the $U(1)$ symmetry in this model is $Z_4$ instead of $Z_2$, which means there can be different types of domain walls and some of them can be potentially unstable \cite{Wu:2022stu,Wu:2022tpe}. 
Therefore, we only make a qualitative description of the picture and leave more detailed studies to the future.

\subsection{GW from first-order phase transition}

As the model includes new scalars and symmetries, new phase transitions can happen when the scalars get VEVs and symmetries are broken.
The most general potential for the new scalar fields allowed by the global $U(1)$ symmetries is  
\begin{eqnarray}
    V_0(\phi_1,\,\phi_2,\,\theta) &=& -\mu_1^2|\phi_1|^2 -\mu_2^2|\phi_2|^2 + \lambda_1|\phi_1|^4 + \lambda_2|\phi_2|^4 + \xi|\phi_1|^2|\phi_2|^2 \nonumber\\
    && - \mu_\theta^2|\theta|^2  + \lambda_\theta|\theta|^4 + \zeta_1|\phi_1|^2|\theta|^2 + \zeta_2|\phi_2|^2|\theta|^2 + (\kappa\phi_1^*\phi_2\theta^2 + \text{h.c.})\,. \label{eq:potential}
\end{eqnarray}
To simplify the discussion, we first consider the terms only evolving $\phi_1$ and $\phi_2$, which is 
\begin{eqnarray}
    \tilde{V}_0(\phi_1,\,\phi_2) = -\mu_1^2|\phi_1|^2 - \mu_2^2|\phi_2|^2 + \lambda_1|\phi_1|^4 + \lambda_2|\phi_2|^4 + \xi|\phi_1|^2|\phi_2|^2 \,. \label{eq:potential_sim}
\end{eqnarray}
The scalar singlet fields $\phi_i$ can be split into the norm fields $\varphi_i$ and the phase fields $\alpha_i$ through $\phi_i = \varphi_i e^{i\alpha_i}/\sqrt{2}$. 
The phase fields do not play any role in the scalar potential.
Thus, the scalar potential can be expressed as 
\begin{eqnarray}
    \tilde{V}_0(\varphi_1,\,\varphi_2) = -\frac{\mu_1^2}{2}\varphi_1^2 - \frac{\mu_2^2}{2}\varphi_2^2 + \frac{\lambda_1}{4}\varphi_1^4 + \frac{\lambda_2}{4}\varphi_2^4 + \frac{\xi}{4}\varphi_1^2\varphi_2^2 \,. \label{eq:potential_var}
\end{eqnarray}

Assuming a hierarchical symmetry breaking pattern  $\langle\varphi_1\rangle\ll\langle\varphi_2\rangle$, the mixing between $\varphi_1$ and $\varphi_2$ can induce a cubic term in the $\varphi_1$ potential after $\varphi_2$ gets a VEV, which enhances the strength of first order phase transition of $\varphi_1$ \cite{Kehayias:2009tn,DiBari:2023mwu}.
The zero-temperature minimisation conditions for the scalar potential are
\begin{eqnarray}
	\mu_1^2 = \lambda_1 \, v_1^2 + \frac{\xi}{2} v_2^2\,,\quad
	\mu_2^2 = \lambda_2 \, v_2^2 + \frac{\xi}{2} v_1^2\,,
\end{eqnarray}
where $v_1 = \langle\varphi_2\rangle$ and $v_2 = \langle\varphi_1\rangle$.
As $\varphi_1$ and $\varphi_2$ get VEVs, the corresponding Higgs fields $\delta{\varphi}_1 = \varphi_1 - v_1$ and $\delta{\varphi}_2 = \varphi_2 - v_2$ develops a mass mixing. 
After $\varphi_2$ gets a VEV, the scalar fields can be expressed in terms of the new mass eigenstates $\overline{\varphi}_1$ and $\overline{\varphi}_2$ as
\begin{eqnarray}
\varphi_1 = \bar{\varphi}_1 \cos{\vartheta} - \delta\bar{\varphi}_2\sin{\vartheta} \label{varphi1}\,,\quad
\varphi_2 = v_2 + \bar{\varphi}_1 \sin{\vartheta} + \delta\bar{\varphi}_2\cos{\vartheta}\label{varphi3}.
\end{eqnarray}
In the hierarchical limit $v_1\ll v_2$, the mixing angle $\vartheta$ can be estimated as 
\begin{eqnarray} \label{e1}
	\theta \simeq -\frac{\xi v_1}{2\lambda_2 v_2}. 
\end{eqnarray}
Such a mixing leads to a cubic term for $\overline{\varphi}_1$ from quartic interaction $\varphi_1^2\varphi_2^2$, which reads
\begin{eqnarray} \label{e2}
	\frac{\xi}{4}\varphi_1^2 \varphi_2^2 \xrightarrow{v_2 \gg v_1} -\frac{\xi^2 v_1}{4\lambda_2} \bar{\varphi}_1^3 + \ldots\,.
\end{eqnarray}
The cubic term enhances the strength of the first order phase transition and makes the corresponding GW detectable.

During a strong first-order phase transition, the GW can be produced mainly through three different mechanisms \cite{Caprini:2024hue}: {\bf collision} between the scalar vacuum bubbles, overlap of the {\bf sound wave} in the plasma and the magnetohydrodynamic (MHD) {\bf turbulence}. 
The total GW spectrum is the sum of the three contributions
\begin{eqnarray}
\Omega_{\rm tot}(f) = \Omega_{\rm bc}(f) + \Omega_{\rm sw}(f) + \Omega_{\rm MHD}(f)\,,
\end{eqnarray}
where bc, sw and MHD stand for bubble collision, sound wave and magnetohydrodynamic turbulence, respectively.
A fourth contribution from particle production from bubble walls have been identified recently \cite{Inomata:2024rkt}, which will not be discussed in detail in this paper.
All three contributions depend on the phase transition dynamics, which can be determined by four key parameters: the wall velocity $\xi_w$, the inverse phase transition duration $\beta/\mathcal{H}_*$, the phase transition strength $\alpha$ and the transition temperature $T_*$. 
Following \cite{Caprini:2024hue}, we adopt the broken power law model for the GW from bubble collision and double broken power laws for the latter two. 
For details, see Appendix \ref{app:GW_FOPT}.

\begin{figure}
    \centering
    \subfloat[][benchmark point 1: $\alpha=0.29$, $\xi_w=1$, \\$\beta/H_*=244.65$, $T_*=5863.12$ GeV]{\includegraphics[width=0.48\linewidth]{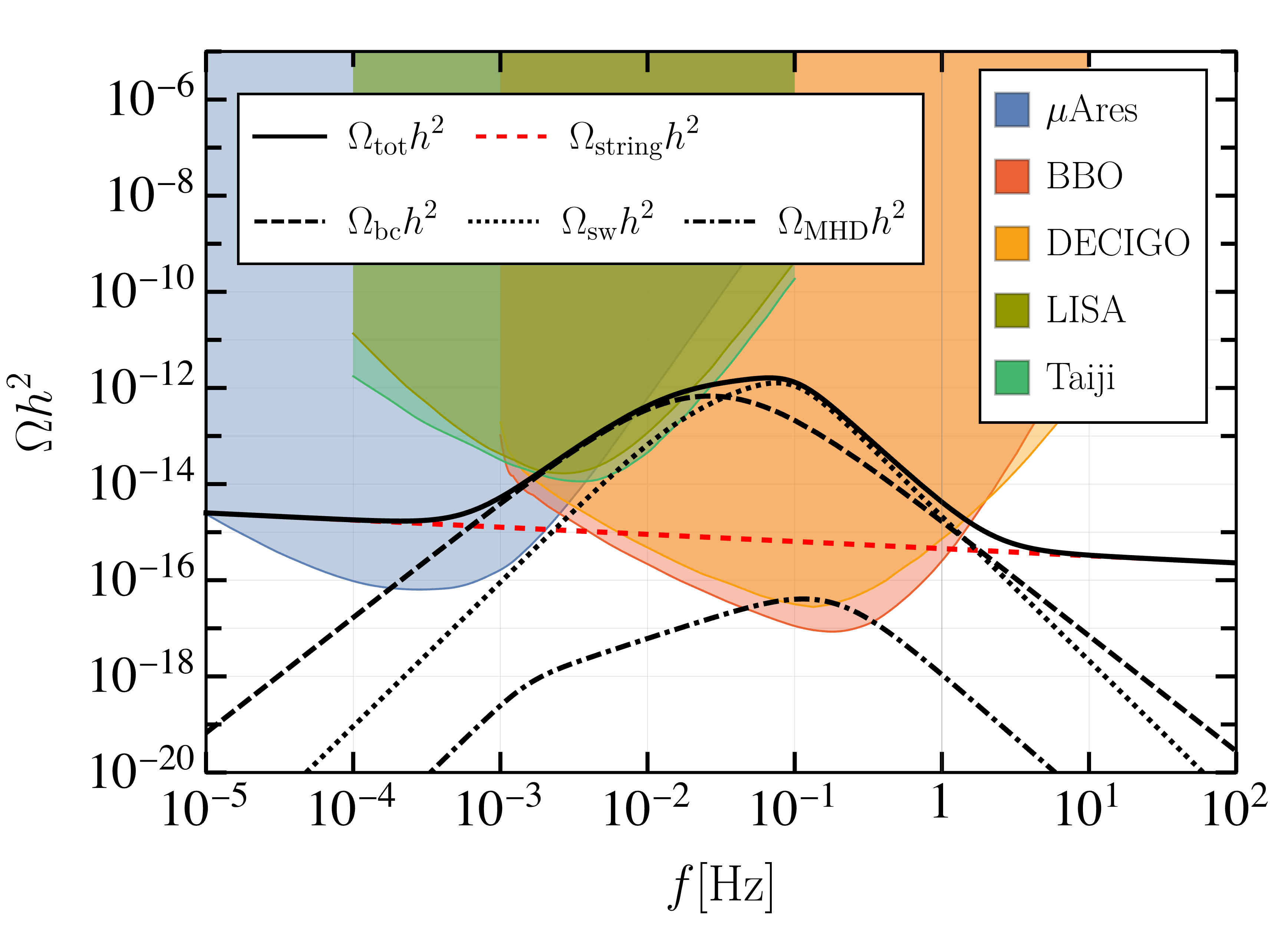}}
    \subfloat[][benchmark point 2: $\alpha=0.30$, $\xi_w=1$, \\$\beta/H_*=204.66$, $T_*=7.8\times10^5$ GeV]{\includegraphics[width=0.48\linewidth]{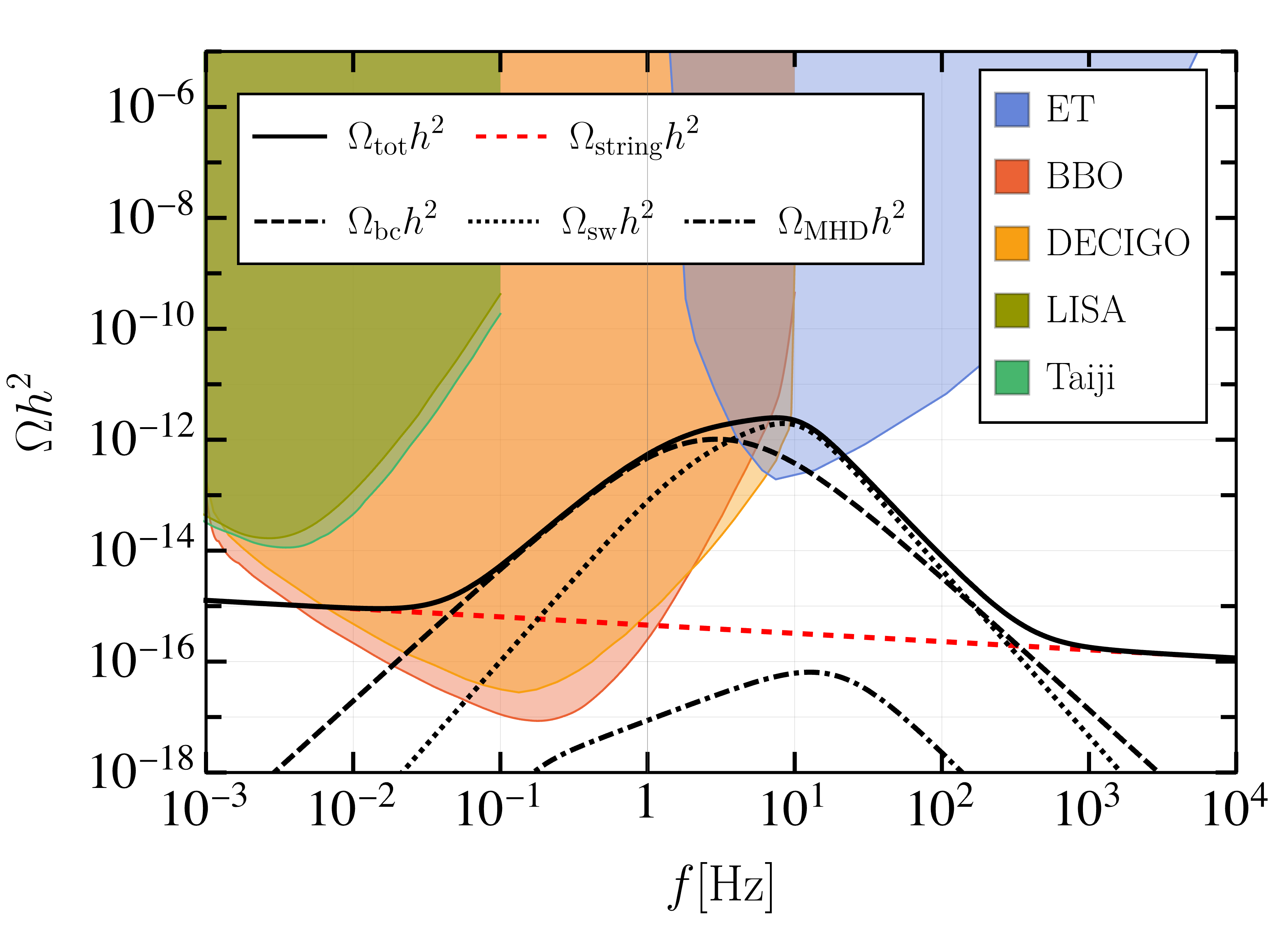}}
    \caption{\it Total GW signal from the first-order phase transition and global cosmic strings. 
    For the first-order phase transition, we choose two benchmark points. 
    For global strings, we choose the $U(1)_{B-L}$ symmetry breaking scale to be $2\times 10^{14}$ GeV for both benchmark points. }
    \label{fig:GW}
\end{figure}

When $\phi_2$ gets a VEV, the global $U(1)_{B-L}$ would be broken and global cosmic strings are formed \cite{Fu:2023nrn}.
The dynamics of the global cosmic string network can also produce primordial GW background \cite{Battye:1993jv,Chang:2019mza,Chang:2021afa}, the signal strength of which relies on the symmetry breaking scale.  
The GW from global cosmic string and first-order phase transition join together, forming a special shape of the GW spectrum.
The total GW signals of two example benchmark points are shown in Fig.\ref{fig:GW} as the solid black lines.
In order to compare the signal with observational sensitivity in different frequency range, we choose $\alpha=0.29$, $\beta/H_*=244.65$, $\xi_w=1$, and $T_*=5863.12$ GeV as benchmark point 1 and $\alpha=0.30$, $\beta/H_*=204.66$, $\xi_w=1$, and $T_*=7.8\times10^5$ GeV as benchmark point 2 for first-order phase transition.
The values of these phase transition parameters can be realised when the VEV of $\phi_1$ is $1.19\times 10^3$ GeV scale and when the VEV of $\phi_1$ is $2.32\times 10^5$ GeV, respectively, according to \cite{DiBari:2023mwu}. 
For global string, we take the $U(1)_{B-L}$ symmetry breaking scale, which is the only free parameter that matters in determining the GW signal, to be $\text{Max}[\langle \phi_1\rangle, \langle \phi_2\rangle] = 2\times 10^{14}$ GeV.
The total GW spectrum is shown in Fig.\ref{fig:GW}.
In order to compare the result with detections, we shadow the region that can be detected by BBO \cite{Corbin:2005ny}, DECIGO \cite{Kawamura:2019jqt,Kawamura:2020pcg}, LISA \cite{Caprini:2015zlo}, Taiji \cite{Chen:2023zkb}, $\mu$Ares \cite{Sesana:2019vho}, and ET \cite{Hild:2008ng} with different colours using power-law sensitivity curves \cite{Thrane:2013oya}.
In benchmark point 1, the transition temperature is TeV scale, and the peak frequency of the corresponding GW spectrum is around decihertz.
As a result, the GW signal can be potentially detected by space-based interferometers. 
In benchmark point 2, the transition temperature is higher, and thus the corresponding GW spectrum peaks at a higher frequency, which provides a target for ground-based interferometers.

Through the GW, it is possible to probe the scale of the heavier Majoron VEV and the lighter Majoron VEV simultaneously, which also control the masses of the two RHNs. 
While the heavier Majoron VEV determines the GW from global cosmic string, the lighter Majoron VEV plays an important role in the GW from the first order phase transition. 
Therefore, we see how elegantly the hierarchical structure between the VEVs and consequently the RHN masses can be probed via GW.

\medskip

\section{Summary and Conclusion \label{sec:con}}

In this paper, we have proposed the first realistic UV complete multi-Majoron model, providing a natural setting for hierarchical right-handed neutrinos. 
As a minimal example, we considered a type-I seesaw model with two complex scalar Majoron fields and two right-handed neutrinos, although the model may be readily extended to more.  
To explain the mass hierarchy of the right-handed neutrinos, two new scalar fields $\phi_1$ and $\phi_2$ are introduced. 
A global $U(1)_N$ symmetry ensures that each right-hand neutrino only couples to one of the complex scalar Majoron fields. As a result, the hierarchy of the right-handed neutrino masses is determined by the hierarchy of the vacuum expectation values of the new complex scalar Majoron fields. 
As the usual interaction between lepton doublets and right-handed neutrinos is forbidden under the $U(1)_N$, a new scalar $\theta$ and two new fermions $\chi_1$ and $\chi_2$ are necessary. 
With the help of the new particles, the Yukawa coupling between the Higgs doublet, lepton doublets and the RHNs can be realised effectively.
The model can be easily extended to include a third RHN, making all the left-handed neutrinos massive. 
The multiple Majoron fields provide a natural explanation for hierarchical RHNs.

Within the minimal multi-Majoron model (MMMM) just described, we have discussed how the neutrino mass and mixing arises from the type I seesaw mechanism. 
As the two RHNs are both coupled to three flavours of lepton doublets, the neutrino mass and mixing can be easily explained, providing the first consistent and realistic model of its kind.
We have discussed how the symmetry breaking results in GWs produced from a combination of global cosmic strings and a first-order phase transitions, leading to a complex network of hybrid defects. This invites a full {\it quantitative} analysis, beyond the scope of the present paper; here we have only provided a {\it qualitative} discussion about the evolution of the network. 
The most remarkable qualitative features are that the GW emission would be enhanced when the string network re-enters the scaling regime after the wall formation, and that the mixing between the scalars enhances the FOPT, which also produces a detectable GW signal in space-based and ground-based interferometers. 
Together with the GWs from global cosmic strings, the total GW spectrum in the MMMM has a novel and  characteristic shape, which provides the tantalising prospect that the mass scales of the hierarchical RHN masses may be empirically measured by GW experiments, providing a quantitative determination of two key parameters of the high scale seesaw mechanism.

\section*{Acknowledgement}
The authors thank Moinul Rahat for collaborating during the initial stages of the project and Maximilian Berbig and Qaisar Shafi for helpful discussions. 
BF acknowledges the Guangdong Basic and Applied Basic Research Foundation No.~2025A1515011079.
SFK thanks CERN for hospitality and acknowledges the STFC Consolidated Grant ST/X000583/1 and the European Union's Horizon 2020 Research and Innovation programme under Marie Sk\l{}odowska-Curie grant agreement HIDDeN European ITN project (H2020-MSCA-ITN-2019//860881-HIDDeN).

\appendix

\section{Gravitational wave from FOPT\label{app:GW_FOPT}}
We follow \cite{Caprini:2024hue} to calculate the spectrum of the GW produced during the first order phase transition.
For consistence, we choose $\{\xi_w,\, \beta/\mathcal{H}_*,\, \alpha,\, T_*\}$ as the input parameters.
For the GW from bubble collision, we adopt the broken power law, which can be expressed as 
\begin{equation}
\Omega_{\mathrm{GW}}^{\mathrm{BPL}}\left(f, \vec{\theta}_{\mathrm{Cosmo}}\right)=\Omega_p \frac{\left(n_1-n_2\right)^{\frac{n_1-n_2}{a_1}}}{\left[-n_2\left(\frac{f}{f_p}\right)^{-\frac{n_1 a_1}{n_1-n_2}}+n_1\left(\frac{f}{f_p}\right)^{-\frac{n_2 a_1}{n_1-n_2}}\right]^{\frac{n_1-n_2}{a_1}}}\,,
\end{equation}
where $f_p$ and $\Omega_p$ are the peak frequency and peak amplitude, respectively.
The parameters $n_1$, $n_2$ and $a_1$ are fixed as $n_1 = 2.4$, $n_2 = -2.4$ and $a_1 = 1.2$ by numerical simulation. 
The peak frequency $f_p$ and peak amplitude $\Omega_p$ are given by  
\begin{equation}
f_p \simeq 0.11 \mathcal{H}_{*, 0} \frac{\beta}{\mathcal{H}_*}\,, \quad h^2 \Omega_p = h^2 F_{\mathrm{GW}, 0} A_{\mathrm{str}} \tilde{K}^2\left(\frac{\mathcal{H}_*}{\beta}\right)^2\,.
\end{equation}
$\tilde{K} \equiv \alpha/(1+\alpha) $ is the fractional energy density of the GW source.
$A_{\mathrm{str}}$ is fixed to be around $0.05$ by the numerical simulations.
$F_{\mathrm{GW}, 0}$ is the red-shift factor for the fractional energy density 
\begin{equation}
h^2 F_{\mathrm{GW}, 0}=h^2\left(\frac{a_*}{a_0}\right)^4\left(\frac{\mathcal{H}_*}{\mathcal{H}_0}\right)^2 \simeq 1.64 \times 10^{-5}\left(\frac{100}{g_*}\right)^{1 / 3} \,.
\end{equation}
The subscript ``$*$'' stands for the value of the quantities at the time of GW production while the subscript ``$0$'' that the value of the quantities today. 
$\mathcal{H}_{*, 0}$ is the Hubble rate at the time of GW production redshifted to today, given by 
\begin{equation}
\mathcal{H}_{*, 0}=\frac{a_*}{a_0} \mathcal{H}_* \simeq 1.65 \times 10^{-5} \mathrm{~Hz}\left(\frac{g_*}{100}\right)^{\frac{1}{6}}\left(\frac{T_*}{100 \mathrm{GeV}}\right)
\end{equation}

For the GW from sound wave and turbulence, we use the double broken power law
\begin{equation}
\Omega_{\mathrm{GW}}^{\mathrm{DBPL}}\left(f, \vec{\theta}_{\text {Cosmo }}\right) = \Omega_2 \times S_2(f)
\end{equation}
where the shape function is given by 
\begin{equation}
S_2(f) = \left(\frac{f}{f_2}\right)^{n_1}
\left[\frac{1+\left(f/f_1\right)^{a_1}}{1+\left(f_2/f_1\right)^{a_1}}\right]^{\frac{-n_1+n_2}{a_1}} 
\left[\frac{1+\left(f/f_2\right)^{a_2}}{2} \right]^{\frac{-n_2+n_3}{a_2}}
\end{equation}
with $f_1$ and $f_2$ two frequency breaks.
$\Omega_2$ is the amplitude at $f_2$. 

In the spectrum of the GW from sound waves, $\Omega_2$ is related to the integrated amplitude $\Omega_{\mathrm{int}}$ by 
\begin{equation}
\Omega_2=\frac{1}{\pi}\left(\sqrt{2}+\frac{2 f_2 / f_1}{1+f_2^2 / f_1^2}\right) \Omega_{\mathrm{int}} \approx 0.55 \Omega_{\mathrm{int}}
\end{equation}
while the integrated amplitude is given by 
\begin{equation}
h^2 \Omega_{\mathrm{int}}=h^2 F_{\mathrm{GW}, 0} A_{\mathrm{sw}} K^2\left(\mathcal{H}_* \tau_{\mathrm{sw}}\right)\left(\mathcal{H}_* R_*\right)\,.
\end{equation}
$A_{\mathrm{sw}}$ determined by simulation to be around 0.11.
$K$ is the kinetic energy fraction given by $K\simeq 0.6\kappa \alpha/(1+\alpha)$, with $\kappa$ the fraction of a single expanding bubble depending on $\alpha$ and $\xi_w$ \cite{Espinosa:2010hh}.
$\tau_{\mathrm{sw}}$ is the duration of the sound wave source, determined by $\tau_{\mathrm{sw}}=\text{min}[\mathcal{H}_*R_*/\sqrt{3K/4},1]$.
$R_*$ is the average bubble size at the time of transition, which can be related to the inverse duration by
\begin{eqnarray}
\mathcal{H}_*R_*\simeq(8\pi)^{1/3}\text{Max}(\xi_w,\,c_s)\frac{\mathcal{H}_*}{\beta}\,.
\end{eqnarray} 
The threshold frequency breaks are 
\begin{equation}
f_1 \simeq 0.2 \mathcal{H}_{*, 0}\left(\mathcal{H}_* R_*\right)^{-1} \quad\text{and}\quad f_2 \simeq 0.5 \mathcal{H}_{*, 0} \Delta_w^{-1}\left(\mathcal{H}_* R_*\right)^{-1}\,,
\end{equation}
where $\Delta_\omega=\xi_\mathrm{shell}/\text{max}(\xi_w,c_s)$, with with $\xi_\mathrm{shell} = |\xi_w - c_s|$ the dimensionless sound shell thickness and $c_s$ the speed of sound.
The parameters $n_1$, $n_2$, $n_3$ and $a_1$, $a_2$ are fixed as $n_1 = 3$, $n_2 = 1$, $n_3 = -3$ and $a_1 = 2$, $a_2 = 4$ by numerical simulation.

For the GW from MHD turbulence, the threshold frequency breaks and the amplitude at $f_2$ are determined by
\begin{equation}
\begin{gathered}
f_1 \simeq \frac{\sqrt{3 \Omega_s}}{4} \mathcal{H}_{*, 0}\left(\mathcal{H}_* R_*\right)^{-1}, \quad f_2 \simeq 2.2 \mathcal{H}_{*, 0}\left(\mathcal{H}_* R_*\right)^{-1}, \\
h^2 \Omega_2=h^2 F_{\mathrm{GW}, 0} A_{\mathrm{MHD}} \Omega_s^2\left(\mathcal{H}_* R_*\right)^2.
\end{gathered}
\end{equation}
$\Omega_s$ is the total energy density fraction defined as $\Omega_s = \varepsilon K$, with $\varepsilon$ the fraction of the kinetic energy that is converted to turbulence.
$\varepsilon $ is chosen to be 0.05 following \cite{Caprini:2019egz}.   
$A_{\mathrm{MHD}}\simeq 4.37 \times 10^{-3}$ is the amplitude parameter for MHD turbulence. 
The parameters $n_1$, $n_2$, $n_3$ and $a_1$, $a_2$ are fixed as $n_1 = 3$, $n_2 = 1$, $n_3 = -8/3$ and $a_1 = 4$, $a_2 = 2.15$ by numerical simulation.

\bibliographystyle{JHEP}
\bibliography{Ref}

\end{document}